\documentclass[aip,jcp,reprint]{revtex4-1}
\usepackage{color}
\usepackage{epsfig}
\usepackage{subfigure}
\usepackage{amsmath}
\usepackage{amsfonts}

\newcommand{\braketop}[3]{\langle #1|#2|#3\rangle}

\begin{document}

\title{Microcanonical Effective Partition Function for the Anharmonic Oscillator}
\date{Submitted on \today}

\author{Jonathan L. Belof}
\email{belof1@llnl.gov}
\affiliation{Lawrence Livermore National Laboratory\\7000 East Avenue., Livermore, CA 94550\\}

\author{Brian Space}
\affiliation{Department of Chemistry, University of South Florida\\4202 E. Fowler Ave., Tampa, FL 33620\\}

\begin{abstract}
The microcanonical effective partition function, constructed from a Feynman-Hibbs potential, is derived using generalized ensemble theory.  The form of the effective Hamiltonian is amenable to Monte Carlo simulation techniques and the relevant Metropolis function is presented.  Using the derived expression for the microcanonical effective partition function, the low-temperature entropy of a proton in an anharmonic potential is numerically evaluated and compared with the exact quantum mechanical canonical result.

\end{abstract}

\maketitle

The Green's function for the quantum dynamic propagator,
\begin{eqnarray}
G(x^\prime, t^\prime; x, t) = \braketop{x^\prime}{e^{-i\hat{H}(t^\prime-t)/\hbar}}{x} \label{eq:greens_function}
\end{eqnarray}
can be expressed in it's path integral form\cite{feynman}, after Trotter factorization and making use of the resolution of the identity, as:\cite{schulman}
\begin{eqnarray}
G(x^\prime, t^\prime; x, t) = \lim_{P \rightarrow \infty} \int\limits_{-\infty}^{\infty} dx_1...dx_{P-1}  \left(\frac{m}{2\pi i \hbar \epsilon}\right)^{P/2} \nonumber \\
\times e^{\frac{i\epsilon}{\hbar} \int\limits_t^{t^\prime} d\tau\, \left\{ \frac{1}{2}m \left( \frac{dx}{d\tau} \right)^2 - V[x(\tau)] \right\} } \label{eq:path_integral_qm}
\end{eqnarray}
where the time interval $\epsilon = (t^\prime - t)/P$ and the path from $x \rightarrow x^\prime$ has been discretized among $P$ points.  Analytically continuing Eq. (\ref{eq:path_integral_qm}) \emph{via} the substitution $\beta = it/\hbar$ and letting the initial time $t = 0$ results in
\begin{eqnarray}
G(x^\prime, -i\beta\hbar; x) = \lim_{P \rightarrow \infty} \int\limits_{-\infty}^{\infty} dx_1...dx_{P-1}  \left(\frac{m}{2\pi \hbar^2 \beta}\right)^{P/2} \nonumber \\
\times e^{-\frac{1}{\hbar} \int\limits_0^{\beta \hbar} d\tau\, \left\{ \frac{1}{2}m \left( \frac{dx}{d\tau} \right)^2 + V[x(\tau)] \right\} } \label{eq:path_integral_statmech}
\end{eqnarray}

It can be shown that the canonical partition function $Q(N,V,\beta$) results from taking the trace of expression (\ref{eq:path_integral_statmech}), where the paths propagate from $x\rightarrow x$:
\begin{eqnarray}
Q(N,V,\beta) = \int\limits_{-\infty}^{\infty} dx\, G(x, -i\beta\hbar; x) \nonumber \\
= \int\limits_{-\infty}^{\infty} dx\, \lim_{P \rightarrow \infty} \int\limits_{-\infty}^{\infty} dx_1...dx_{P-1}  \left(\frac{m}{2\pi \hbar^2 \beta}\right)^{P/2} \nonumber \\
\times e^{-\frac{1}{\hbar} \int\limits_0^{\beta \hbar} d\tau\, \left\{ \frac{1}{2}m \left( \frac{dx}{d\tau} \right)^2 + V[x(\tau)] \right\} } \label{eq:path_integral_partition_function}
\end{eqnarray}
and the integration is done over all possible closed paths that start and end at $x$.  A great deal of applied research has proceeded from the approximation whereby Eq. (\ref{eq:path_integral_partition_function}) is closed for a finite Trotter number $P$.  Indeed, in such a form expression (\ref{eq:path_integral_partition_function}) then looks very much like a classical partition function (albeit with a $\beta$-dependent harmonic term) and can be numerical evaluated by many-dimensional integration techniques such as Monte Carlo.  An alternative to this discretized approach is to write Eq. (\ref{eq:path_integral_partition_function}) as a Fourier expansion around the path and numerically solve for the Fourier coefficients.

However, a third traditional approach is to approximate the integrals over $\int\limits_{-\infty}^{\infty} dx_1...dx_{P-1}$ \emph{via} a variational principle\cite{feynman_hibbs} the result of which is then expressed as an exponential of an \emph{effective potential} $\widetilde{W}(x)$:

\begin{eqnarray}
Q(N,V,\beta) \approx \widetilde{Q}(N,V,\beta) = \int\limits_{-\infty}^{\infty} dx\, \sqrt{\frac{m}{2\pi \hbar^2 \beta}} e^{-\beta \widetilde{W}(x)} \label{eq:Q}
\end{eqnarray}
where $\widetilde{Q}(N,V,\beta)$ is the \emph{canonical effective partition function}.  At low-temperature we hope to capture the quantum effects present in Eq. (\ref{eq:path_integral_partition_function}) (to what degree depends crucially on $\widetilde{W}$) and in the high-temperature limit $\widetilde{Q}$ is equivalent to the classical partition function.  In it's original formulation,\cite{feynman_hibbs} it can be shown that $\widetilde{W}(x)$ satisfies a variational principle if taken to be a Gaussian-smeared potential,

\begin{eqnarray}
\widetilde{W}(x) = \int\limits_{-\infty}^{\infty} dy\, \frac{1}{\sqrt{2\pi a^2}} e^{\frac{(x-y)^2}{2a^2}} U(y) \label{eq:W}
\end{eqnarray}
where in it's first approximation the Gaussian width $a^2 = \beta \hbar^2/12m$ and the same fixed-width approximation is made here.  Techniques that improve upon the fixed-width approximation have been previously made.\cite{feynman_kleinert,kleinert,cowley}  The Taylor series expansion of expression (\ref{eq:W}) yields the familiar terms commonly used in molecular simulation.

The canonical effective partition function has been of significant value in numerical statistical mechanics since it includes quantum fluctuations while preserving the readily understood mathematical structure of the classical partition function.  In the simplest approximation of a fixed-Gaussian width of $\beta \hbar^2/12m$ the effective approach provides accuracy amenable to the semiclassical regime.

Curiously, while the path integral expression for the microcanonical partition function has been derived\cite{freeman,lawson}, to our knowledge the microcanonical effective partition function has not been reported in the literature.  While the canonical ensemble is quite natural for the study of many physical systems, there are cases where the microcanonical ensemble is more convenient since the thermodynamic energy can be fixed.  Our intention here is to develop the microcanonical effective partition function such that quantum fluctuations may be included in, for instance, $NVE$ Monte Carlo\cite{creutz,ray} simulations.

We derive the microcanonical effective partition function through application of generalized ensemble theory\cite{sack,guggenheim,tiller} which allows us to relate the constant energy shell ensembles \emph{directly} to other ensembles in which thermal energy can flow between system and bath.  Of immediate interest is the Laplace transform relationship between the canonical and microcanonical partition functions; this relation between ensembles has been long utilized in semiclassical theory, however to our knowledge it has not been employed in the context of effective potentials.  We begin by demonstrating the following example:

\begin{eqnarray}
Q(N,V,\beta) = \int\limits_0^{\infty} dE\, e^{-\beta E} \Omega(N,V,E)
\end{eqnarray}
since the energy spectrum may always be shifted such that the lower bound is zero.  In solving for the microcanonical partition function, the inverse Laplace transform yields:

\begin{eqnarray}
\Omega(N,V,E) = \frac{1}{2\pi i} \oint d\beta\, e^{\beta E} Q(N,V,\beta) \nonumber \\
= \int\limits_{-\infty}^{\infty} d\Gamma\, \frac{1}{2\pi i} \int\limits_{\gamma - i\infty}^{\gamma + i\infty} d\beta\, e^{\beta(E - H)}
\end{eqnarray}
where $d\Gamma$ is the phase space differential form $(N! \, h)^{-1} dx\, dp$.  Since $\beta = \sigma + i\tau$ and no singularity is present in the right-half of the complex plane, we may take the contour vertically through $\gamma = 0$.  Since $\rm Re(\beta) = 0$ along the integration we may make the substitution $\beta = -i\tau$:

\begin{eqnarray}
\Omega(N,V,E) = \int\limits_{-\infty}^{\infty} d\Gamma\, \frac{1}{2\pi} \int\limits_{-\infty}^{\infty} d\tau\, e^{i\tau(H - E)} \nonumber \\
= \int\limits_{-\infty}^{\infty} d\Gamma\, \delta(H - E) \label{eq:delta_function}
\end{eqnarray} 
which is the microcanonical partition function, as it should be.  Another simple example is the quantum harmonic oscillator:

\begin{eqnarray}
\Omega_{HO} &=& \frac{1}{2\pi i} \oint d\beta\, e^{\beta E} Q_{HO} = \frac{1}{2\pi} \int\limits_{-\infty}^{\infty} d\tau\, e^{-i\tau E} \frac{e^{\frac{1}{2}i\tau \hbar \omega}}{1 - e^{i \tau \hbar \omega} } \nonumber \\
&=& \frac{1}{2\pi} \int\limits_{-\infty}^{\infty} d\tau\, e^{i\tau\left(\frac{1}{2} \hbar \omega - E\right)} \sum_n e^{i\tau \hbar \omega n} \nonumber \\
&=& \sum_n \delta \left[\hbar \omega \left( n + \frac{1}{2}  \right) - E \right] \label{eq:nve_qho}
\end{eqnarray}
where we note that in this case we have integrated over the quantum mechanical partition function, resulting in a discrete series over the eigenspectrum rather than a phase space integral.

Along similar lines, we wish to construct the microcanonical effective partition function $\widetilde{\Omega}(N,V,E)$ from $\widetilde{Q}(N,V,\beta)$ through use of the same Laplace structure.  Proceeding in this manner,

\begin{eqnarray}
\widetilde{\Omega}(N,V,E) = \frac{1}{2\pi i} \oint d\beta\, e^{\beta E} \widetilde{Q}(N,V,\beta) \nonumber \\
= \frac{1}{2\pi i} \oint d\beta\, e^{\beta E} \int\limits_{-\infty}^{\infty} dx\, \sqrt{\frac{m}{2\pi \hbar^2 \beta}} e^{-\beta \widetilde{W}} \nonumber \\
\end{eqnarray}
Using Eq. (\ref{eq:W}), the Gaussian smeared version of the anharmonic potential $U_{AHO} = \frac{1}{2}kx^2 + \frac{1}{4}gx^4$ is exactly integrable and yields
\begin{eqnarray}
\widetilde{W}_{AHO} = U_{AHO} + \frac{\beta \hbar^2}{24m}\left(k + 3gx^2\right) + {\left(\frac{\beta \hbar^2}{24m}\right)}^2 3g \\
= U_{AHO} + W_1(\beta) \label{eq:aho_smeared}
\end{eqnarray}
for the effective potential.  With the momentum integration having been undone, the inverse Laplace transform now becomes:
\begin{eqnarray}
\widetilde{\Omega}_{AHO}(E) = \int\limits_{-\infty}^{\infty} d\Gamma\, \frac{1}{2\pi i} \oint d\beta\, e^{\beta \left[E - H - W_1\left(\beta\right)\right]} \nonumber \\
= \int\limits_{-\infty}^{\infty} d\Gamma\, \frac{1}{2\pi i} \oint d\beta\, e^{\beta\left(E - H\right)} e^{-\beta^2 \left[ \frac{\hbar^2}{24m}\left( k + 3gx^2 \right) \right] } e^{-\beta^3 \left(\frac{\hbar^2}{24m}\right)^2 3g } \nonumber \\
= \int\limits_{-\infty}^{\infty} d\Gamma\, \frac{1}{2\pi i} \oint d\beta\, e^{a\beta - b\beta^2 -c\beta^3}
\end{eqnarray}
where the contour integral can be approximated by the method of steepest descent to yield
\begin{eqnarray}
= \frac{1}{\sqrt{2\pi}} \int\limits_{-\infty}^{\infty} d\Gamma\, \frac{ e^{a\beta_0 -b\beta_0^2 -c\beta_0^3} }{\left( 2b + 6c\beta_0 \right)^{\frac{1}{2}} } \label{eq:sd_aho_nve_fh}
\end{eqnarray}
where the saddle point, $\beta_0$,
\begin{eqnarray}
\beta_0 = \frac{2b - \sqrt{4b^2 + 12ac}}{6c} \label{eq:sad_point}
\end{eqnarray}

Recalling that the following substitutions have been made,
\begin{eqnarray}
a = E - H \nonumber \\
b = \frac{\hbar^2}{24m} \left( k + 3gx^2 \right) \nonumber \\
c = \left(\frac{\hbar^2}{24m}\right)^2 3g \nonumber
\end{eqnarray}
we arrive at the final expression for the microcanonical effective partition function for the anharmonic system
\begin{eqnarray}
\widetilde{\Omega}_{AHO}(E) = \int\limits_{-\infty}^{\infty} d\Gamma\, \frac{e^{-\frac{3\left(H - E\right)^2 }{4b} - \frac{c\left( H - E\right)^3}{8b^3}}}{\sqrt{2\pi\left( 2b - \frac{6ac}{2b} \right)}} \label{eq:aho_nve_fh}
\end{eqnarray}

\begin{figure}[htp]
\includegraphics[width=3.3 in]{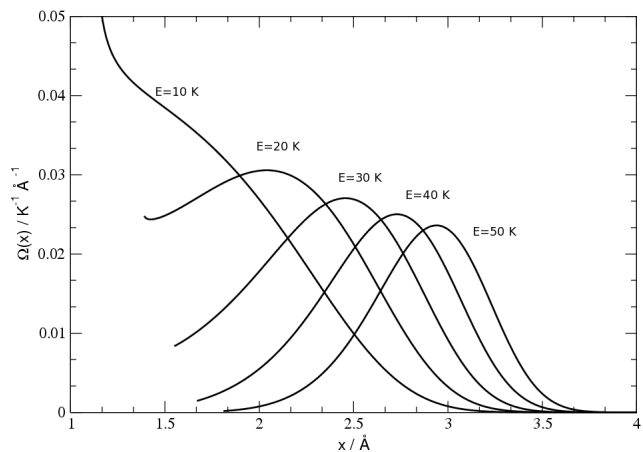}
\caption{The configurational distribution function, $\Omega(x)$, for $k=1$, $g=2$.}
\label{fig:omega_dist}
\end{figure}

We may note some interesting features of the microcanonical effective partition function for this anharmonic system.  The phase space distribution function is dominated by a Gaussian distribution in $H-E$, the Gaussian width being determined by the quantum mechanical factor $b$. In the classical limit of vanishing $b,c$ the Gaussian distribution narrows to approach the classical microcanonical distribution $\delta(H-E)$.  Shown in Fig. \ref{fig:omega_dist} is the $k=1$ and $g=2$ distribution function $\Omega(x)$ (the momentum having been integrated) for several energy values corresponding to the low temperature regime.

\begin{figure}[htp]
\subfigure[$g=4$]{
	\includegraphics[width=3.3 in]{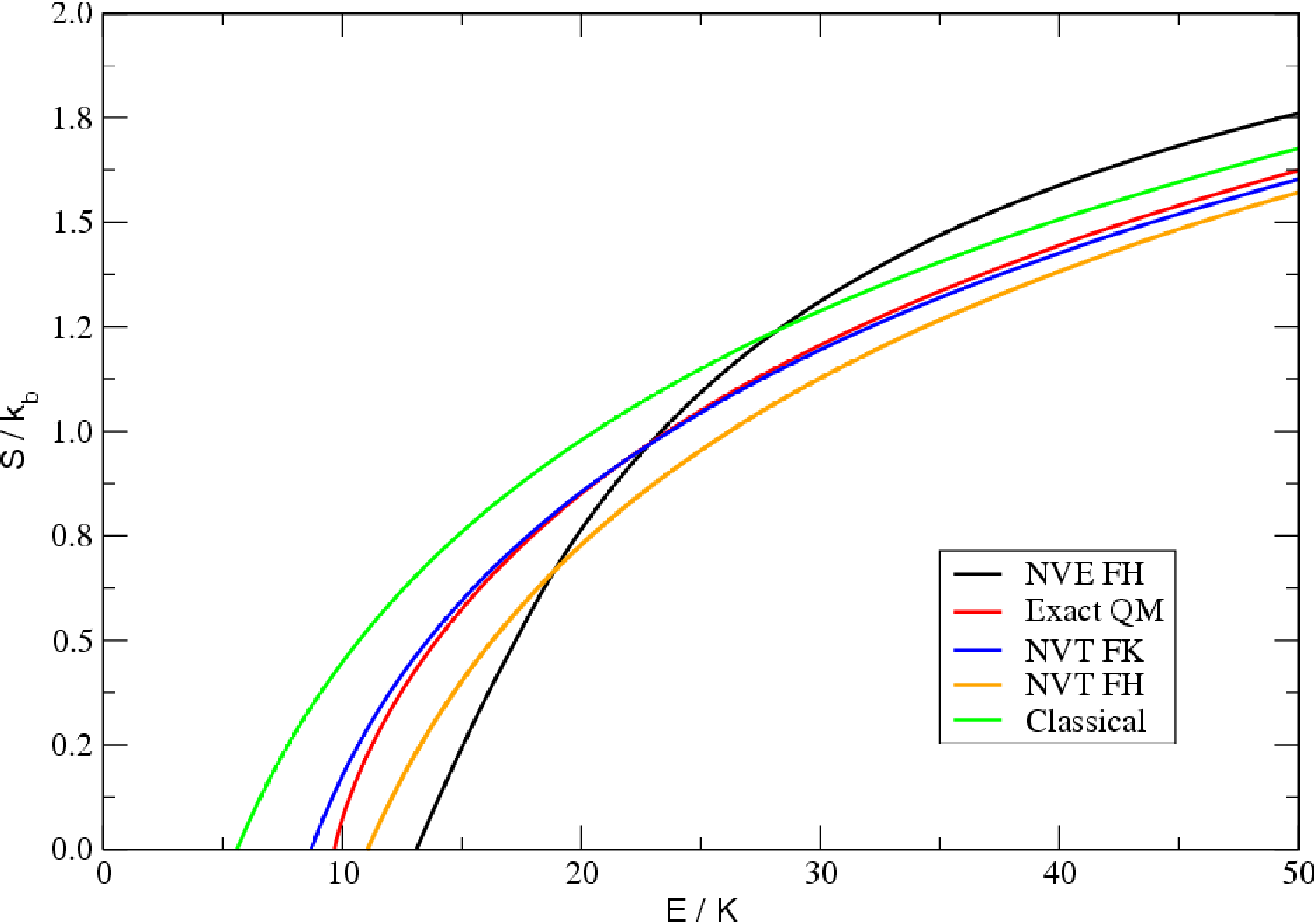}
	\label{subfig:entropy_figure_g4}
}
\subfigure[$g=40$]{
	\includegraphics[width=3.3 in]{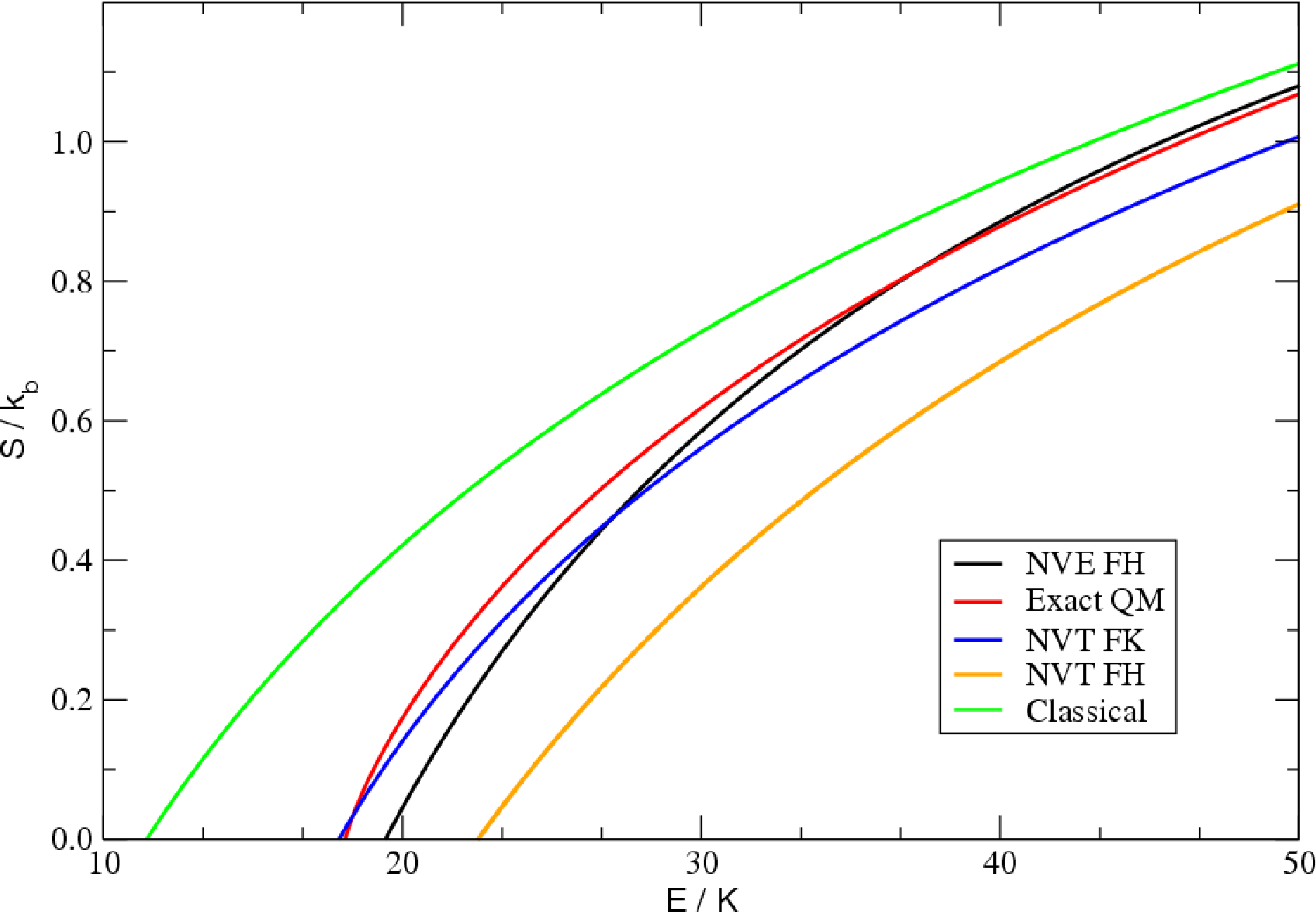}
	\label{subfig:entropy_figure_g40}
}
\subfigure[$g=200$]{
	\includegraphics[width=3.3 in]{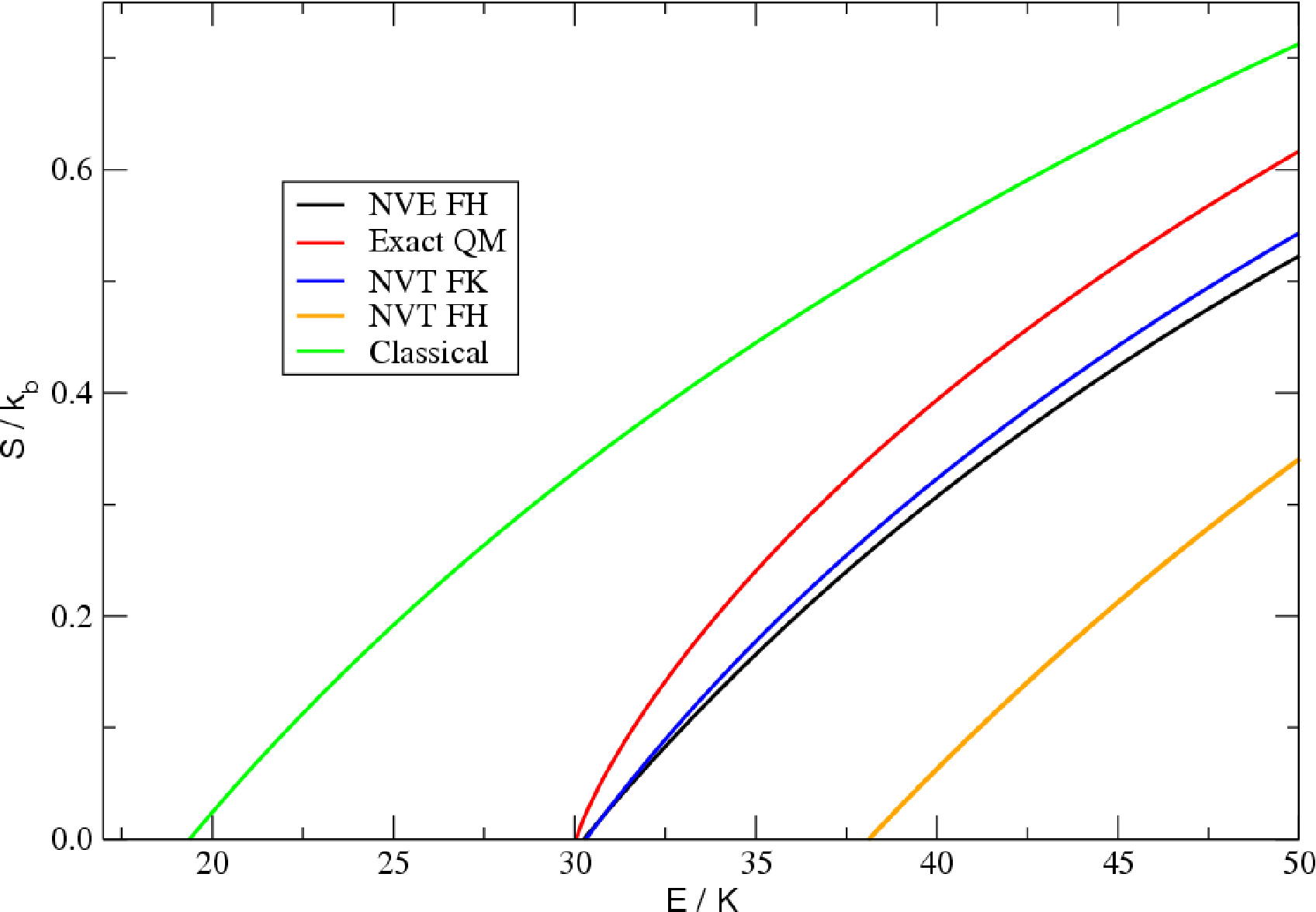}
	\label{subfig:entropy_figure_g200}
}
\caption{The entropy of a proton in an anharmonic well with $k=1$ and $g=4,40,200$ calculated from the microcanonical effective partition function given by Eq. (\ref{eq:aho_nve_fh}) (black), in comparison with the canonical entropy calculated from direct summation of the eigenspectrum (red), the canonical entropy obtained \emph{via} the higher-order effective potential method of Ref. \onlinecite{feynman_kleinert} (blue), the canonical Feynman-Hibbs method of Ref. (\onlinecite{feynman_hibbs}) (orange) and the classical result (green).}
\label{fig:entropy_figure}
\end{figure}

Numerical evalution of Eq. (\ref{eq:aho_nve_fh}) has been performed to obtain the entropy of a proton in an anharmonic well for $k=1$ and $g=4,40,200$.  These values for the anharmonicity of the potential have been chosen since they are among those available for the calculated eigenspectrum of Ref. (\onlinecite{montroll}) and were also used by Feynman and Kleinert in Ref. (\onlinecite{feynman_kleinert}) to illustrate the improvements possible by extending the original fixed-width Gaussian formalism.  We calculate the cumulative entropy $\Sigma$ by integrating Eq. (\ref{eq:aho_nve_fh}):
\begin{eqnarray}
\Sigma(E) = \int\limits_{0}^{E} dE^\prime\, \Omega(E^\prime) \label{eq:cumulative_sigma} \\
S = k \ln \Sigma(E) \label{eq:cumulative_entropy}
\end{eqnarray}
where we note that, in the thermodynamic limit, the microcanonical entropy $k \ln \Omega(E)$, the level density entropy $k \ln \omega(E)$ and the cumulative entropy $\ln k \Sigma(E)$ are all equivalent to within an additive constant.\cite{huang}  However, as was done in Ref. \onlinecite{feynman_kleinert}, a comparison will be made with the canonical partition function for the exact quantum mechanical system, and the discrete sum over levels presented by the canonical partition function of $N=1$ anharmonic oscillators necessitates comparison with the cumulative entropy.

In Fig. \ref{fig:entropy_figure}, the Feynman-Hibbs microcanonical entropy, denoted ``NVE FH'', is compared with the exact quantum mechanical entropy found from the canonical ensemble, $S = k \ln Q + E/T$.  The quantum mechanical $Q$ has been calculated by direct summation of the Boltzmann factors for the first 9 levels\cite{montroll} of the eigenspectrum.  Also shown for comparison is the canonical entropy found \emph{via} the nearly exact method of Feynman and Kleinert\cite{feynman_kleinert} (in this method the width of the Gaussian smear is not held fixed) denoted ``NVT FK'', the standard canonical Feynman-Hibbs method of Ref. (\onlinecite{feynman_hibbs}) (where the Gaussian width is held fixed, as it is also in the current work) denoted ``NVT FH'', and the classical canonical entropy.

With respect to the accuracy of the method, we note several features from Fig. \ref{fig:entropy_figure}.  First, we point out that it is well known that the higher-order FK method will reproduce the exact quantum mechanical result with near perfect accuracy even at very low temperature and for strong anharmonicity -- especially where the more standard NVT FH method will fail.\cite{feynman_kleinert}  In contrast, the method presented here can be seen as the microcanonical analog of the less accurate NVT FH method.  Interestingly, however, this method agrees quite well with both the exact quantum and NVT FK results at low temperature and even for the strong anharmonicity value of $g=200$ (where the commonly used NVT FH method fails) and yet yields poor agreement at higher temperatures and also for the relatively harmonic $g=4$.  This appears to be due to the fact that the integral expressed in Eq. (\ref{eq:aho_nve_fh}) is only well sampled when the energy distribution is broadened (\emph{i.e.} under non-classical conditions) since it becomes increasingly difficult to sample the distribution approaching a delta function with high accuracy -- as the distribution narrows to the Dirac delta function the saddle point approximation breaks down.

Expression (\ref{eq:aho_nve_fh}) may also be derived for various other intermolecular potentials, and can be evaluated by multidimensional phase space integration techniques such as Hybrid Monte Carlo.\cite{duane,mehlig}  The Hybrid Monte Carlo (HMC) technique makes use of an $NVE$ molecular dynamics integrator, with a large non-energy conserving timestep, to sample the phase space integral \emph{via} a Metropolis accept/reject scheme based upon the full Hamiltonian.  Unlike a traditional canonical Monte Carlo scheme where only the configurational part of the integral is sampled, in HMC the momentum integration is performed explicitly by the algorithm through a randomized resampling of the momenta from an equilibrium distribution -- a useful aspect for sampling Eq. (\ref{eq:aho_nve_fh}) given that the momentum dependence in this equation cannot be analytically integrated out.  Such methodology may prove practical for sampling the microcanonical ensemble for atomic and molecular systems, where the HMC algorithm would proceed with a modified Metropolis function based upon Eq. (\ref{eq:aho_nve_fh}):

\begin{eqnarray}
\frac{Pr(i\rightarrow j)}{Pr(j\rightarrow i)} = e^{-\left( \epsilon_j - \epsilon_i \right)} \frac{\zeta_i}{\zeta_j}\label{eq:hmc_metropolis}
\end{eqnarray}
where
\begin{eqnarray}
\epsilon_i = \frac{3}{4b_i} \left(H_i - E \right)^2 + \frac{c}{8b_i^3} \left(H_i - E \right)^3  \nonumber \\
\zeta_i = \sqrt{2\pi\left( 2b_i - \frac{6a_i c}{2b_i} \right) } \nonumber
\end{eqnarray}

The authors acknowledge funding from the U.S. Department of Energy, Basic Energy Sciences (Grant No. DE0GG02-07ER46470).  Lawrence Livermore National Laboratory is operated by Lawrence Livermore National Security, LLC, for the U.S. Department of Energy, National Nuclear Security Administration under Contract DE-AC52-07NA27344.

\end{document}